\documentclass[12pt,sort&compress]{iopart}
\usepackage{xcolor}
\usepackage{graphicx}
\usepackage[hidelinks]{hyperref}

\usepackage[numbers]{natbib}
\begin{document}
\title[Manipulating a beam of barium fluoride molecules using an electrostatic hexapole]{Manipulating a beam of barium fluoride molecules using an electrostatic hexapole}

\author{A~Touwen$^{1,2}$, J~W~F~van Hofslot$^{1,2}$, T~Qualm$^{1,2}$, R~Borchers$^{1,2}$, R~Bause$^{1,2}$, H~L~Bethlem$^{1,3}$, A~Boeschoten$^{1,2}$, A~Borschevsky$^{1,2}$, T~H~Fikkers$^{1,2}$, S~Hoekstra$^{1,2}$, K~Jungmann$^{1,2}$, V~R~Marshall$^{1,2}$, T~B~Meijknecht$^{1,2}$, M~C~Mooij$^{1,3}$, R~G~E~Timmermans$^{1,2}$, W~Ubachs$^{3}$ and L~Willmann$^{1,2}$ 
\\ NL-\emph{e}EDM collaboration}

\address{$^1$ Van Swinderen Institute for Particle Physics and Gravity, University of Groningen, The Netherlands}
\address{$^2$ Nikhef, National Institute for Subatomic Physics, The Netherlands}
\address{$^3$ Department of Physics and Astronomy, LaserLaB, Vrije Universiteit Amsterdam, The Netherlands}
\ead{H.L.Bethlem@vu.nl}

\date{\today}
\begin{abstract}
An electrostatic hexapole lens is used to manipulate the transverse properties of a beam of barium fluoride molecules from a cryogenic buffer gas source. The spatial distribution of the beam is measured by recording state-selective laser-induced fluorescence on an \textsc{emccd} camera, providing insight into the intensity and transverse position spread of the molecular beam. Although the high mass and unfavorable Stark shift of barium fluoride pose a considerable challenge, the number of molecules in the low-field seeking component of the $N = 1$ state that pass a 4\,mm diameter aperture 712\,mm behind the source is increased by a factor of 12. Furthermore, it is demonstrated that the molecular beam can be displaced by up to $\pm$5\,mm by moving the hexapole lens. Our measurements agree well with numerical trajectory simulations. We discuss how electrostatic lenses may be used to increase the sensitivity of beam experiments such as the search for the electric dipole moment of the electron.  

     
\end{abstract}


\maketitle
\section{Introduction} 
Heavy polar molecules are uniquely suitable as probes for symmetry violation beyond what is incorporated in the Standard Model of particle physics\,\cite{Safronova2018}. Currently, the most precise limits on the electric dipole moment of the electron (\textit{e}EDM) are set by experiments on HfF$^{+}$\,\cite{Roussy2022}, ThO\,\cite{Andreev2018} and YbF\,\cite{Hudson2002}. A non-zero value of the \textit{e}EDM would constitute a violation of time-reversal\,(\emph{T})-symmetry and \emph{CP}-symmetry, where \emph{C} denotes charge conjugation and \emph{P} parity. The current bounds for the \textit{e}EDM already put tight constraints on theories that extend the Standard Model\,\cite{Ema2022, Cirigliano2016}. \\ \indent
The NL-\textit{e}EDM collaboration aims to measure the \textit{e}EDM in a slow and cold beam of barium fluoride (BaF) molecules, obtained by Stark deceleration and laser-cooling of an intense cryogenic buffer gas cooled beam\,\cite{Aggarwal2018}. Here, we present experiments on manipulating a beam of BaF molecules using an electrostatic hexapole lens.\\ \indent 
The main motivation for this work is to increase the number of molecules that enters our traveling-wave Stark decelerator. In previous experiments on deceleration of SrF molecules\,\cite{Aggarwal2021}, there was a 370\,mm gap between the cryogenic source and the decelerator to allow for separation of the source and decelerator chambers by a vacuum valve\footnote[1]{Such a valve is necessary, as the source is heated up every night and the target is replaced every few weeks. Breaking the vacuum of the decelerator chamber requires high voltage conditioning of the electrodes which typically takes many hours.}, resulting in a transverse loss of molecules. This loss can be avoided by mapping the emittance of the beam at the exit of the source onto the \emph{transverse} acceptance of the decelerator using an electrostatic lens\,\cite{Meerakker2012}. By mounting the lens on a manipulator, the alignment of the beam into the decelerator can be optimised. In principle, the \emph{longitudinal} phase-space matching of the beam with the acceptance of the decelerator is also improved by increasing the distance between the beam source and decelerator\,\cite{Fabrikant2014}, but for realistic lengths of the lens this increase is limited. \\
\indent Electrostatic multipole lenses were first introduced in the 1960s by Gordon\,\emph{et al.}\,\cite{Gordon1954} and Bennewitz\,\emph{et al.}\,\cite{Bennewitz1955} to perform microwave spectroscopy on state-selected molecular beams and are used for collision studies between state-selected oriented molecules\,\cite{Stolte1991, Meerakker2012, BrouardParkerMeerakker2014}. Many studies have targeted relatively light and simple molecules such as OH, CH, CH$_3$F and NH$_3$ in low-field seeking states that have a large electric dipole moment to mass ratio and a Stark shift that is either predominantly linear or quadratic. In contrast, BaF is both much heavier and has a much less favourable Stark shift over the relevant range of electric field strength, changing from quadratic into linear and ultimately becoming high-field seeking. Nevertheless, it is shown here that the beam intensity down stream from the source can be increased significantly by using a properly designed hexapole lens. The hexapole lens was developed for BaF molecules in the first excited rotational level ($N=1$) of the electronic and vibrational ground state and has also been tested with molecules in the second excited rotational state ($N = 2$) state, that may be used as an upgrade to the experiment in the future. In a related study, Wu \emph{et al.} used a hexapole lens to enhance the molecular flux of ThO molecules for the ACME experiment\,\cite{Wu2022}. 


\begin{figure*}[t]
    \centering
    \includegraphics[width=.9\linewidth]{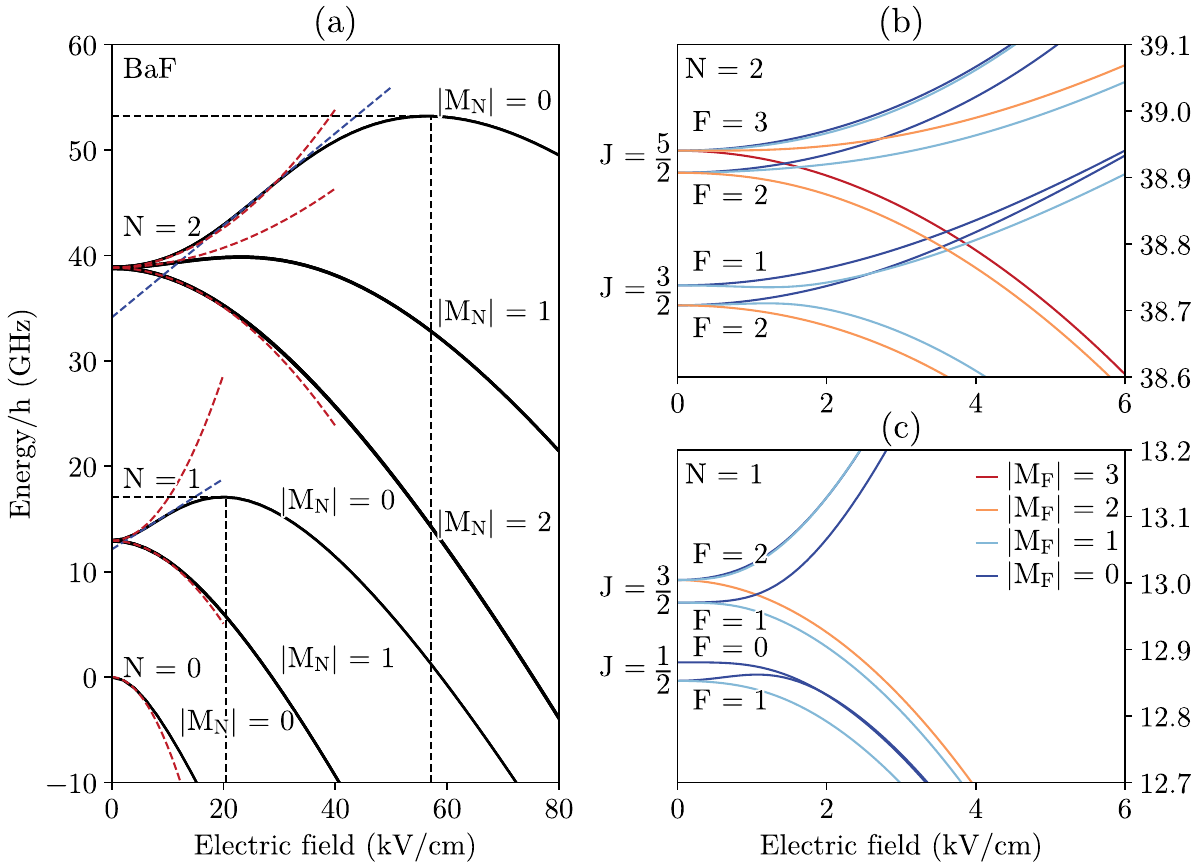}
    \caption{(a) Stark effect in \textsuperscript{138}Ba\textsuperscript{19}F for the lowest rotational states ($N \leq$ 2) of $X^2\Sigma^+ \,v = 0$ calculated using \textsc{pgopher}\,\cite{Western2017} based on molecular constants from Ryslewicz\,\emph{et al.}\,\cite{Ryzlewicz1980} and Ernst\,\emph{et al.}\,\cite{Ernst1986}. The red and blue dashed lines show the quadratic extrapolation at zero and a linear extrapolation at intermediate field strengths to illustrate Stark curves ideal for quadrupole and hexapole lenses, respectively. The dashed black lines indicate the electric field strength and energy when these low-field seeking components become high-field seeking. (b) and (c) show the low electric field regime for $N = 2$ and $N = 1$ respectively, demonstrating their hyperfine structure. 
    }
    \label{fig:Stark}
\end{figure*}

\section{The force on barium fluoride in an electrostatic multipole lens}\label{sec:Force and Stark}
The electrostatic potential $\Phi$ for a charge free region of space fulfills Laplace's equation  $\nabla^2 \Phi = 0$. In the case where it is constant along an axis $z$ it can be written as a multipole expansion in cylindrical coordinates as\,\cite{Meerakker2012},
\begin{equation}
    \Phi (r, \theta, z) 
    = \Phi_0\sum_{n=1}^\infty a_n\left(\frac{r}{r_0}\right)^n\cos(n\theta+\theta_n).
\end{equation}
The field is parameterised by voltage magnitude $\Phi_0$, coefficient $a_n$, inner radius $r_0$ and offset angle $\theta_n$. When considering a single multipole term the resulting electric field strength is,
\begin{equation}
    E_{(n)}(r, \theta, z) = |-\vec{\nabla}\Phi_n| = n\Phi_0\frac{r^{n-1}}{r_0^n},
    \label{Eq:Ehex}
\end{equation}
representing the dipole, quadrupole and hexapole fields for $n = 1, 2, 3$, respectively. \\ \indent
The radial force on a polar molecule within an electric field $E$ is given by $F_r = -\partial W(E)/\partial r$, with $W(E)$ being the potential energy of a molecule in an electric field, also known as the Stark shift. For an ideal lens, this restoring force is linear, $F_r = -kr$, where $k$ is the spring constant of the system.\\ \indent
For a molecule in a state with predominantly quadratic Stark effect $W(E) = \frac{1}{2}\alpha E^2$, such as NH$_{3}$ in the upper inversion level of the $J=1,K=1$ state, a linear restoring force can be established by a quadrupole field $E_{(2)}(r) = 2\Phi_0\,r/r_0^2$ \,\cite{Gordon1954,Cheng2016}. In a state with a linear Stark shift $W(E) = - \mu_{\mathrm{\textit{eff}}} E$, such as OH in the upper level of the $\Lambda$-doublet in the $J=1,\Omega=3/2$ state\,\cite{Schreel1993} or ND$_{3}$ in the upper inversion level of the $J=1,K=1$ state\,\cite{Butkovskaya1976}, a linear restoring force can be established with a hexapole field $E_{(3)}(r) = 3\Phi_0\,r^2/r_0^3$ instead. \\
\indent Let us now turn to barium fluoride. The solid lines in \autoref{fig:Stark} show the Stark effect of the lowest rotational states of BaF in the lowest vibrational state ($v=0$) of the electronic ground state ($X^2\Sigma^+$) found by diagonalisation of the Stark Hamiltonian using \textsc{pgopher}\,\cite{Western2017}. Here the electric dipole moment, $\mu_e = 3.1702(15)$ D, and the hyperfine constants are taken from Ernst\,\emph{et al.}\,\cite{Ernst1986} and other molecular constants, including the rotational constant, $B = 6.47395465(11)$\,GHz, from Ryzlewicz\,\emph{et al.}\,\cite{Ryzlewicz1980}. \\ \indent
The rotational levels, labeled by $N$, are split due to the spin-rotation interaction with the spin of the molecules, ($S = 1/2$), due to the unpaired valence electron, resulting in fine structure states labeled by $J$. These fine structure states are further split due to spin-spin and spin-rotation interaction with the nuclear spin of the fluorine ($I = 1/2$) atom, resulting in four hyperfine states labeled by the total angular momentum $F$. The Stark shift of these hyperfine states is shown in \autoref{fig:Stark}(b) and (c). At higher electric field, different $M_F$ levels group together and shift collectively depending on $M_N$. 
\indent To describe the motion of the molecules within the multipole lens, we will neglect the hyperfine splittings and only consider the shift of the rotational level $N, |M_N|$ as a whole. \\
\indent At low field strength, the Stark shift of these levels can be well approximated by perturbation of the Stark Hamiltonian up to second-order\,\cite{Brown2003}, as,
\begin{equation}
    W_{lf}(E) = W^{(0)}(N) + W^{(2)}(N, M_N)\frac{\mu_e^2E^2}{B}.
\end{equation}
The odd power terms vanish (for a negligible \textit{e}EDM) because of the odd parity of the field\,\cite{Cohen1984}. This quadratic expansion is shown as red dashed lines in \autoref{fig:Stark}(a). Note the splitting into low and high-field seeking states, curving up and down for increasing field strength. \\ \indent
When the electric field is increased further the contribution of higher rotational levels increases and the low-field seeking Stark curves follow a linear trend. The blue dashed lines in \autoref{fig:Stark}(a) show a linear fit to the low-field seeking Stark curves around the maximum slope which happens at a field of 7.1\,kV/cm for the $N = 1$ state and 25.4\,kV/cm for the $N = 2$ state. The effective dipole moments, $\mu_{\mathrm{\textit{eff}}} = -\partial W(E)/\partial E$, at these points are found to be $-0.665$\,D and $-0.867$\,D for the $|M_N| = 0$ substates of $N = 1$ and $N = 2$, respectively.\\ \indent
At even higher electric field strength, the Stark shift becomes comparable to the rotational splittings and the rotational states become high-field seeking. The states with $|M_N| = 0$, which start as low-field seeking at low electric field strengths, turn into high-field seeking at turning point field strengths $E_{tp}$ of $20.4$\,kV/cm and $57.1$\,kV/cm for $N = 1$ and $N = 2$, respectively. \\ \indent
The energy shift at these turning point fields provide the maximum potential energy for molecules in these states in an electric field. From relating this to kinetic energy, the resulting maximum transverse capture velocity of any electrostatic multipole lens for the heavy BaF molecules ($m$ = 157\,amu) in these states is equal to $4.6$\,m/s and $8.6$\,m/s, corresponding to a maximum trap depth of $32$\,mK and $110$\,mK, respectively. 
\section{Design of the hexapole lens}
Due to the shape of the low-field seeking Stark shift described above, molecules in a quadrupole field experience a linear restoring force in a regime of low field close to the molecular beam axis and thereby perform a harmonic oscillation. Molecules in higher fields further away from the axis will experience a reduced restoring force and will oscillate with a lower frequency that depends on the amplitude of their motion. In principle, a perfect lens can be constructed by using a large quadrupole and low voltage, however, the capture velocity of such a lens would be very small, making it ineffective. \\
\indent As the orifice of the source and the spatial acceptance of the decelerator are both significant compared to a typical size of a multipole lens, some image distortion is acceptable for our purpose. To maximise the mapping into the acceptance region the ideal lensing for intermediate displacements is more important than for molecules close to the axis. Therefore, we have chosen to use a hexapole lens.\\
\indent To maximise the velocity of the molecules that will be refocused by the lens, the electric field strength at the electrodes should be equal to the turning point field $E_{(n)}(r_0) = E_{tp}$. A higher maximum electric field strength would reduce the effective hexapole diameter. For a hexapole with inner radius of 6\,mm, this optimal field strength for molecules in the $N = 1, |M_N| = 0$ is achieved at a voltage magnitude of 4\,kV. 
\begin{figure}[ht]
    \centering
    \includegraphics[width=.6\textwidth]{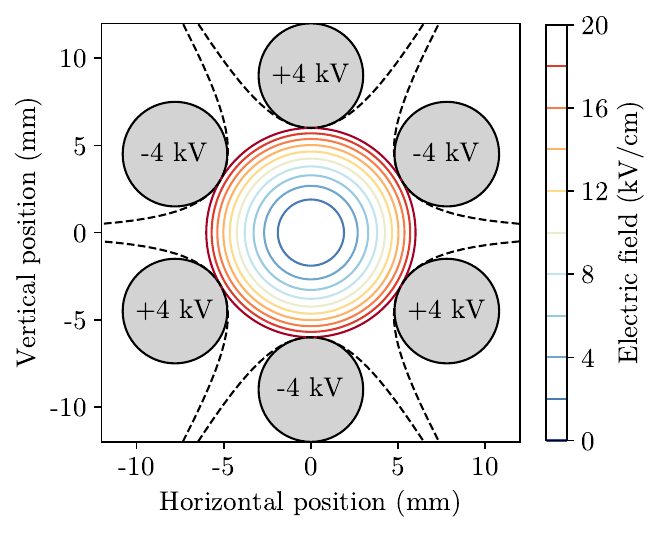}
    \caption{Electric field in an ideal hexapole with inner radius of $r_0$ = 6\,mm and applied voltage magnitude of $\Phi_0$ = 4\,kV. The ideal hyperbolic electrodes are shown as black dashed lines, the grey circles show the practical implementation using cylindrical electrodes with a 6\,mm diameter.}
    \label{fig:hexapole field}
\end{figure}

\indent \autoref{fig:hexapole field} shows the electric field that is created by a hexapole formed by six hyperbolic electrodes to which alternating voltages of plus or minus 4\,kV are applied. 
For ease of manufacturing, we use cylindrical rods which have a diameter of 6\,mm and are placed with a minimal distance to the center of 6\,mm. These rods are indicated by the grey circles in the figure. The use of cylindrical electrodes instead of hyperbolic electrodes introduces small deviations from the ideal field\,\cite{Meerakker2012} with limited impact on our results. \\ \indent
An analytic description of the dynamics of particles in a hexapole lens system and the relations to properties known from optics can be found in Cho\,\emph{et al.}\,\cite{Cho1991}. To estimate the length scale of the system, the oscillation length can be considered,
\begin{equation}\label{eq:oscillation length}
    l = 2\pi\frac{v_l}{\omega} = 2\pi\,v_l\sqrt{-\frac{m\,r_0^3}{6\mu_{\mathrm{\textit{eff}}}\Phi_0}},
\end{equation}
where $v_l$ is the longitudinal velocity of the molecules. Here the linear fit to the Stark curve at maximum slope is used as was described before. For a molecule in the $N = 1, |M_N| = 0$ state in a hexapole with inner radius $r_0$ of $6$\,mm and applied voltage $\Phi_0$ of $4$\,kV and longitudinal velocity of $184$\,m/s this corresponds to a half oscillation length of $l/2 = 594$\,mm.  In the lower and higher field regimes the restoring force is smaller, resulting in a longer effective oscillation length. \\
\indent The non-linearity of the Stark curve results in a spread of the molecular beam at the focus point of the hexapole lens, analogous to spherical aberrations in optical systems. These aberrations can in principle be compensated for by introducing a sequence of quadrupole and hexapole lenses as demonstrated by Everdij \emph{et al.}\,\cite{Everdij1973}, but this comes at a cost of increased complexity and reduced flexibility.
The focus is further softened by the large velocity spread of our cryogenic buffer gas cooled molecular beam, analogous to chromatic aberrations in optical systems. Chromatic aberrations can be reduced by applying pulsed voltages to the hexapole rods as discussed by Ke \emph{et al.}\,\cite{Ke2016} (see also Crompvoets \emph{et al.}\,\cite{Crompvoets2005}).
\begin{figure}[ht]
    \centering
    \includegraphics[width=.6\textwidth]{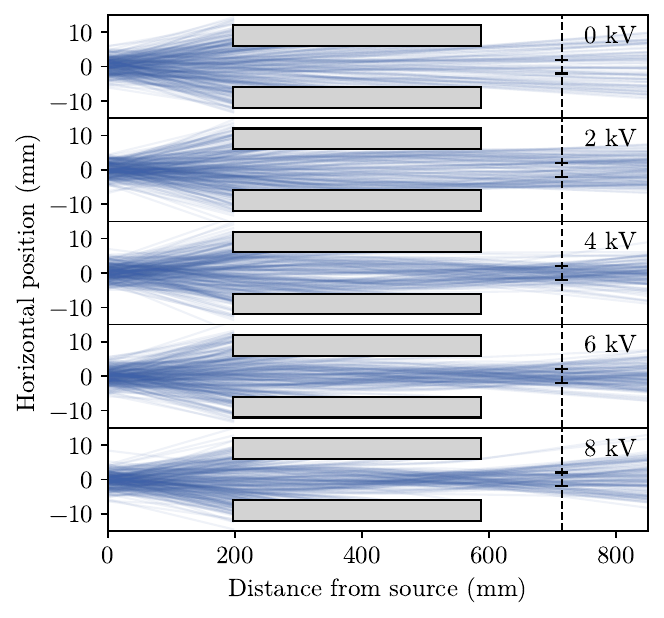}
    \caption{Simulated trajectories of barium fluoride molecules in the $N = 1, |M_N| = 0$ state for five different voltages applied to the hexapole electrodes as indicated. The position of the hexapole lens is indicated by the grey rectangles. The dashed black line shows the position of the position of the detection zone aimed for in this experiment. The markings show to the target region of 4\,mm corresponding to the inner diameter of the decelerator. The initial distribution of the molecular beam is chosen to be only in the shown transverse direction for clarity. The transverse velocity spread is limited as molecules with higher velocity do not make it to the target region.}
    \label{fig:Trajectories}
\end{figure}

\indent To optimise the design and understand the behaviour of the hexapole lens a Monte Carlo trajectory simulation package has been developed. \autoref{fig:Trajectories} shows simulated trajectories for different applied voltage magnitudes to the hexapole lens, demonstrating the oscillation length of the system and the transverse spread of the molecular beam at the detection zone.\\ 
\indent In the simulated trajectories the previously described motion of the molecules through the lens can be seen. As understood from (\ref{eq:oscillation length}), the focal length decreases for increasing voltage. For voltages above 4\,kV, molecules close to the electrodes experience an electric field that is beyond the turning point of the Stark curve and are being defocused. As a result, the effective inner diameter of the hexapole lens is reduced.

\begin{figure}[ht]
    \centering
    \includegraphics[width=.65\textwidth]{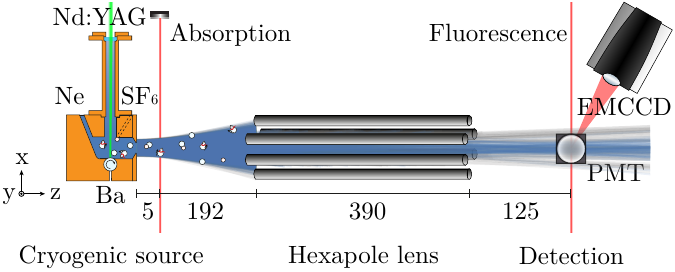}
    \caption{Experimental setup as seen from above with molecules traversing from left to right. Longitudinal distances are given in mm (not to scale). The pulsed molecular beam is produced in the cryogenic source, monitored with absorption, focused transversely by the hexapole lens and imaged using laser-induced fluorescence on a \textsc{pmt} above the molecular beam and on an \textsc{emccd} camera at a 60 degree angle from the molecular beam. The grey and blue lines show trajectories for molecules traversing the hexapole at an applied voltage of 0 and 4\,kV, respectively.
    }
    \label{fig:setup}
\end{figure}
\begin{figure}[ht]
    \centering
    \includegraphics[width=.65\textwidth]{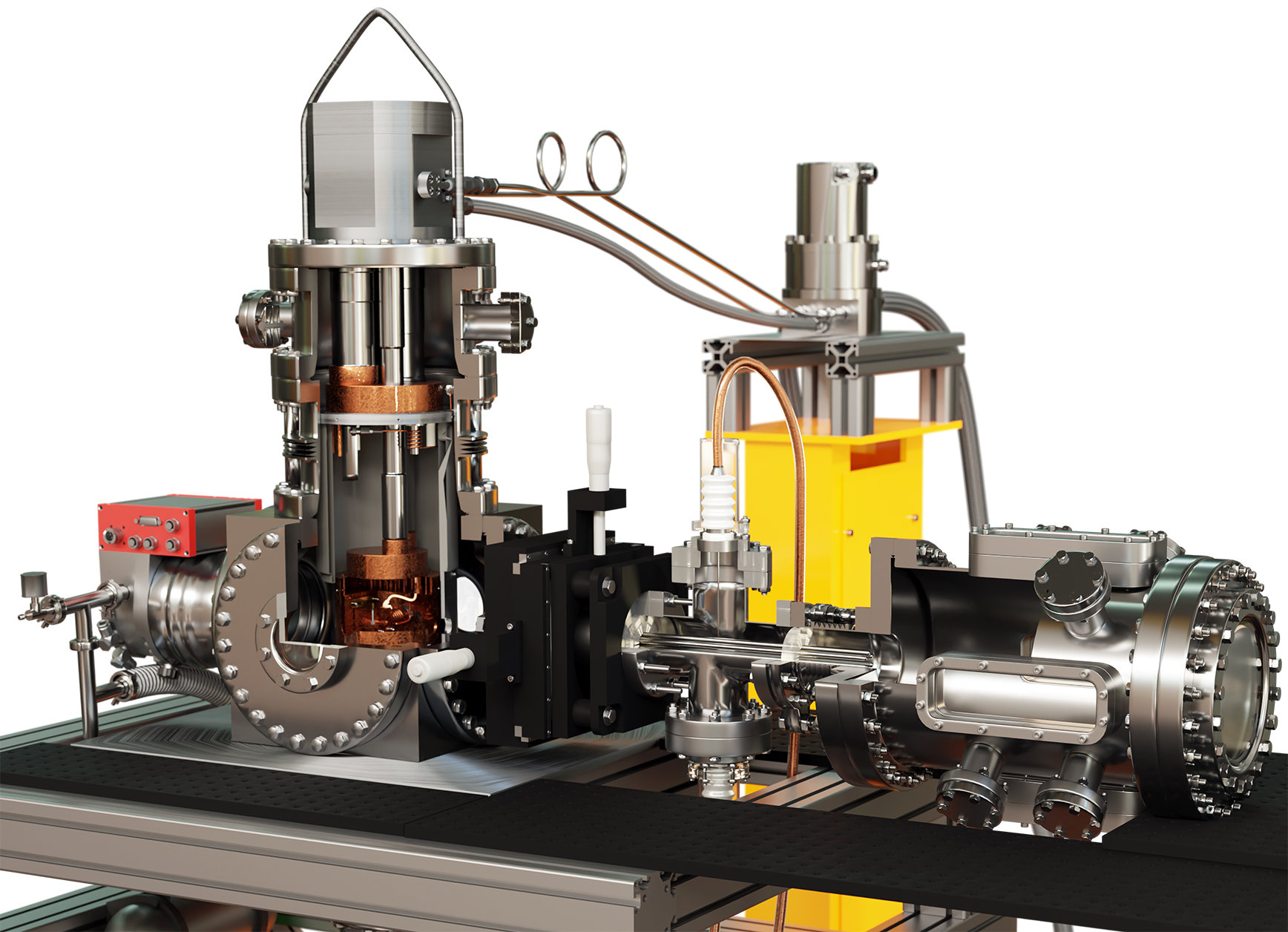}
    \caption{Render of the implementation of the setup described in \autoref{fig:setup} with partial removal of vacuum chambers. The tube that holds the hexapole lens shown in the middle can be displaced using a manipulator, shown in black. The chamber used for fluorescence detection has been designed to enable transverse laser-cooling of the molecular beam close to the hexapole lens exit.}
    \label{fig:setup render}
\end{figure}
\begin{figure*}[t]
    \centering
    \includegraphics[width=0.8\linewidth]{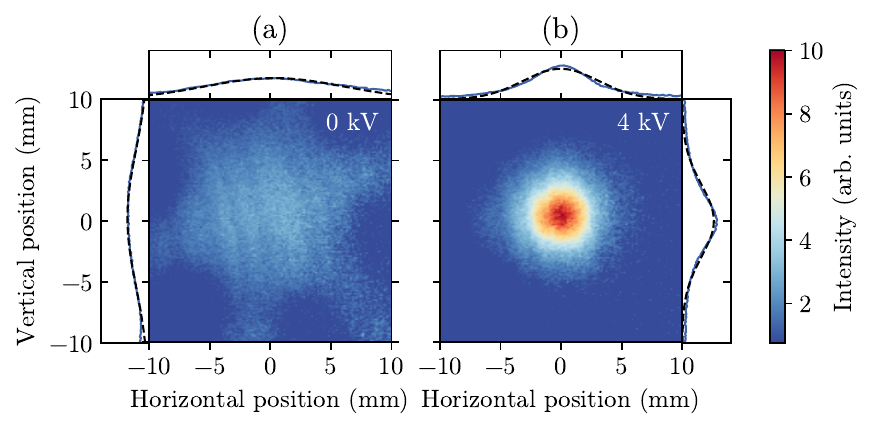}
    \caption{Images of the cross section of the molecular beam at the detection zone (a) without voltage and (b) with near optimal voltage applied to the hexapole lens (4\,kV). The molecules are imaged by laser-induced fluorescence captured on an \textsc{emccd} camera, probing molecules in the $N = 1, J = 3/2$ ground state. The projections along each orthogonal axis (blue curves) are shown together with Gaussian fits (black dashed-line). In (a) the free-flight of the molecular beam is only obstructed by the electrodes, resulting in the six shadows. From these shadows it follows that the hexapole is mounted at a $\sim$10\,degree angle compared to \autoref{fig:hexapole field}.}
    \label{fig:Axis projection}
\end{figure*}
\section{Methods}\label{sec:methods}
An overview of the setup consisting of a cryogenic source, hexapole lens and detection zone is shown in \autoref{fig:setup}. A render of the practical implementation of the setup can be found in \autoref{fig:setup render}. The pulsed beam of barium fluoride molecules is produced in the cryogenic source, in a copper cell cooled to 17\,K. A rotating Ba metal target is ablated with a 5\,ns pulse of 3-5\,mJ in a 1\,mm diameter focus from an Nd:YAG laser at a wavelength of 532\,nm. The ablation product forms BaF molecules in a chemical process involving SF\textsubscript{6}\,gas. The molecules are cooled by collisions with a neon buffer gas and expand through the 4.5\,mm diameter cell exit to form a molecular beam.\\
\begin{figure}[ht]
    \centering
    \includegraphics[width=.55\textwidth]{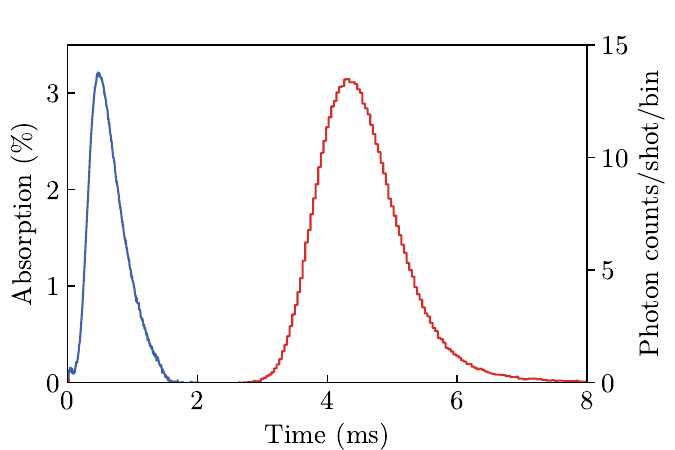}
    \caption{Time of flight profiles of the pulsed molecular beam at 5\,and 712\,mm from the cell exit measured by absorption (blue curve) and fluorescence (red curve) detection respectively. The double-pass absorption percentage is averaged over 40 $\mu$s, the fluorescence photon arrival times are presented in 40 $\mu$s bins. In total this corresponds to $\sim$500\,photons per shot captured on the \textsc{pmt}. 
    }
    \label{fig:Time of flight}
\end{figure}
\indent The molecular beam is monitored by absorption detection at 5\,mm from the cell exit by passing through a retro-reflected 860\,nm laserbeam probing the $X^2\Sigma^+ \,v = 0, \,N = 1$ or $N = 2$ to $A^2\Pi_{1/2} \,v' = 0, \,J' = 3/2$ or $J' = 5/2$ transitions. The returning laserlight is captured on a photo-diode (Thorlabs PDB210A) and digitised on an analog-to-digital converter (Picoscope 5444D) with a time resolution of 400\,ns. A typical time of flight profile is shown as the blue curve in \autoref{fig:Time of flight}. The laserlight for detection is produced by a Ti:Sapphire laser (MSquared SolsTiS PSX-XF) pumped by a diode‐pumped solid‐state laser (Lighthouse Photonics Sprout-D) and locked to a wavelength meter (HighFinesse WS8-2). The wavelengthmeter data is also read out and stored with the molecule signal to be used in spectroscopy measurements. \\
\indent After a total free-flight of 197\,mm the molecules enter a hexapole lens with an inner diameter of 12\,mm, consisting of six cylindrical rods of 6\,mm diameter and 390\,mm length connected in an alternating pattern to a positive and negative high voltage supply, as is shown \autoref{fig:hexapole field}. \\
\indent After another 125\,mm of free-flight the molecular beam intersects the perpendicular probe laser sheet and laser-induced fluorescence is detected at the same transition as absorption. The laser sheet has a 22\,mm height with flat-top intensity profile, and a 0.3\,mm Gaussian full width at half maximum (\textsc{fwhm}). For time sensitive measurements, the fluorescence light is captured on a photomultiplier tube (\textsc{pmt}, Hamamatsu HS7421-50) mounted orthogonally to both the molecular beam axis and the laser beam. The \textsc{pmt} has an effective solid angle $\Omega/4\pi = 2.2 \times 10^{-3}$ and quantum efficiency of 8.5\%. Single photon pulses captured on the \textsc{pmt} are extended and recorded on the same analog-to-digital converter as used for absorption. Pulse detection software records the arrival time of these photons with a 400\,ns time resolution resulting in a time of flight profile as shown as the red curve in \autoref{fig:Time of flight}. From these time sensitive measurements the longitudinal velocity is deduced to be 184\,m/s with a Gaussian \textsc{fwhm} spread of 64\,m/s, which is used as an input for the simulations. \\
\indent Fluorescence light is also captured on an electro-multiplying charged-coupled device (\textsc{emccd} camera, Andor iXon DU897) in the plane of the molecular and laser beam at an angle of 60\,degrees from the molecular beam axis. This angle was chosen to avoid the background radiation originating from the afterglow of the barium target, which could not be sufficiently suppressed by using narrow-band filters (Semrock FF01-860/11-25). The angle is compensated for by scaling the horizontal axis in post processing. A slight trapezoidal distortion of the image is neglected. To verify this simplification the horizontal and vertical projections are considered separately in the analysis presented in this paper showing equivalent results. \\
\indent To capture a signal image on the camera, a time window from 2 to 8\,ms after the ablation pulse is used, as this is when the the molecules traverse the laser sheet as shown in the time of flight in \autoref{fig:Time of flight}.
To compensate for the significant readout background counts in the camera, a background image is taken when no molecules are present and subtracted from the signal image. The timing is incorporated in the measurement sequence controlled by the delay generator (Berkeley Nucleonics Model 575). Typically an image presented here consists of the average over 1000\,shots, which corresponds to 100\,seconds of data taking. An example of a molecular beam image is shown in \autoref{fig:Axis projection}.
\section{Spectra with and without hexapole lens}\label{sec:spectra}
\begin{figure}[ht]
    \centering
    \includegraphics[width=.65\textwidth]{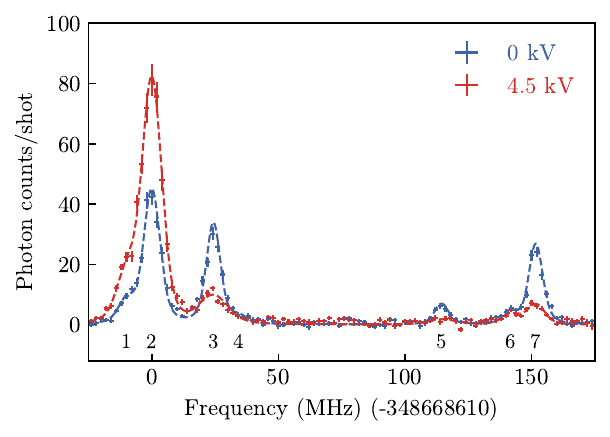}
    \caption{Spectra of the $X^2\Sigma^+ \,v = 0, \,N = 1$ to $A^2\Pi_{1/2} \,v' = 0, \,J' = 3/2$ transition at a binsize of 1\,MHz with the hexapole lens off (blue datapoints) and at a voltage magnitude near optimal focus (red datapoints). The dashed blue and red lines result from a fit using multiple Voigt profiles. Seven transitions are expected in this frequency range, which are labeled for ease of reference.}
    \label{fig:Spectrum N=1}
\end{figure}
\indent In \autoref{fig:Spectrum N=1}, spectra recorded using the \textsc{pmt} are shown of the transitions from the $X^2\Sigma^+ \,v = 0, \,N = 1$ to the $A^2\Pi_{1/2} \,v' = 0, \,J' = 3/2$ with the hexapole lens off (blue data points) and at a voltage magnitude near the optimal focus (red data points). In these measurements, the laser power in the sheet is 35\,$\mu$W and the linewidths of the observed transitions are limited by residual Doppler broadening. Four main peaks, corresponding to transitions originating from the four hyperfine components in the ground state, are easily distinguishable. Three of these peaks are further split due to hyperfine structure in the excited state. The dashed blue and red lines are the result of a fit using multiple Voigt profiles. The position and intensity of the lines correspond well to those predicted by \textsc{pgopher} using the molecular constants from Ernst\,\emph{et al.}\,\cite{Ernst1986}, Ryzlewicz\,\emph{et al.}\,\cite{Ryzlewicz1980}, Steimle\,\emph{et al.}\,\cite{Steimle2011} and Marshall\,\emph{et al.}\cite{Marshall2023}. The transitions are labelled 1 to 7 for ease of reference. The zero point of the frequency axis is set at the $X^2\Sigma^+ \,v = 0, \,N = 1, J=3/2, F=2$ to the $A^2\Pi_{1/2} \,v' = 0, \,J' = 3/2, F'=2$ transition at 348668610\,MHz\footnote[2]{The offset of these spectra are determined based on \textsc{pgopher} simulations of the molecule based on literature and the absolute frequencies from Mooij~\emph{et al.}~\cite{mooij2024novel}}, labelled 2 in the figure.

\indent When a voltage is applied to the hexapole, molecules in a low-field seeking state are focused into the detection zone, while molecules in a high-field seeking state are defocused. As a result, the intensity increases or decreases based on the distribution of low- and high-field seeking components of the ground state that is probed in the transition.  


We will now discuss the effect of the hexapole lens on the different transitions. Transitions 5 to 7 originate from the $N=1, J=1/2, F=0,1$ states which are high-field seeking. Consequently, when voltage is applied to the hexapole, the intensity of these transitions decrease as expected. Transition 1 probes the $|M_F|=0,1$ components of the $N=1, J=3/2, F=2$ which are low-field seeking. Consequently, when voltage is applied to the hexapole, the intensity increases. Transition 2, 3 and 4 probe a mix of low-field and high-field seeking states. The intensities depends strongly on the angle between polarisation of the light and the magnetic field.

\indent Besides being more intense, the width (\textsc{fwhm}) of transition 1 and 2 is also increased, from 6.1\,MHz to 7.7\,MHz, as a result of the increased velocity spread of the focused molecules. The observed linewidths correspond to velocity spreads of 5.2\,m/s and 6.6\,m/s, respectively.
\begin{figure}[ht]
    \centering
    \includegraphics[width=.65\textwidth]{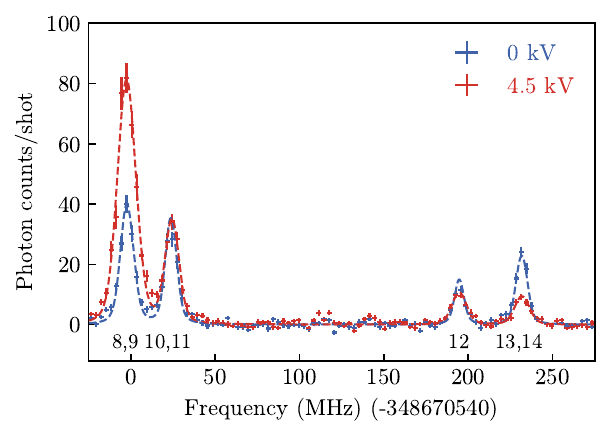}
    \caption{Spectra of the $X^2\Sigma^+ \,v = 0, \,N = 2$ to $A^2\Pi_{1/2} \,v' = 0, \,J' = 5/2$ transition at a binsize of 2\,MHz with the hexapole lens off (blue datapoints) and at a voltage magnitude near the optimal focus (red datapoints). The dashed blue and red lines result from a fit using multiple Voigt profiles. Seven transitions are expected in this frequency range, which are labeled 8 to 14 for ease of reference. The hyperfine structure in the excited state is unresolved in our measurement. The unassigned peaks correspond to transitions in \textsuperscript{137}Ba\textsuperscript{19}F.}
    \label{fig:Spectrum N=2}
\end{figure}

In \autoref{fig:Spectrum N=2} similar spectra are shown of the transitions from the $X^2\Sigma^+ \,v = 0, \,N = 2$ to $A^2\Pi_{1/2} \,v' = 0, \,J' = 5/2$, both with the hexapole lens off (blue data points) and at a voltage magnitude near the optimal focus (red data points). Again four main peaks, corresponding to transitions originating from the four hyperfine components in the ground state, are easily distinguishable. The hyperfine structure in the excited state is unresolved in this measurement. The transitions are labelled 8 to 14 for ease of reference. The zero point of the frequency axis is set at the $X^2\Sigma^+ \,v = 0, \,N = 2, J=5/2, F=2$ to the $A^2\Pi_{1/2} \,v' = 0, \,J' = 5/2, F=3$ transition at 348670640\,MHz\footnotemark[2], labelled 9 in the figure.\\
\indent The $N = 2$ state contains components that increase in energy ($|M_N| = 0$), stay approximately constant ($|M_N| = 1$) or decrease in energy ($|M_N| = 2$) when the electric field is increased. Transitions 8 and 9 originate mainly from low-field seeking states, transitions 10-12 originate mainly from $|M_N| = 1$ states, while transitions 13 and 14 mainly target high-field seeking states. Indeed, when applying voltages to the hexapole, peaks 8 and 9 are broadened and become more intense compared to the situation when no voltage is applied. The intensity of peaks 13 and 14 is reduced when voltage is applied, while the intensity of 10-12 remains approximately the same. 
\begin{figure}[ht]
    \centering
    \includegraphics[width=.6\textwidth]{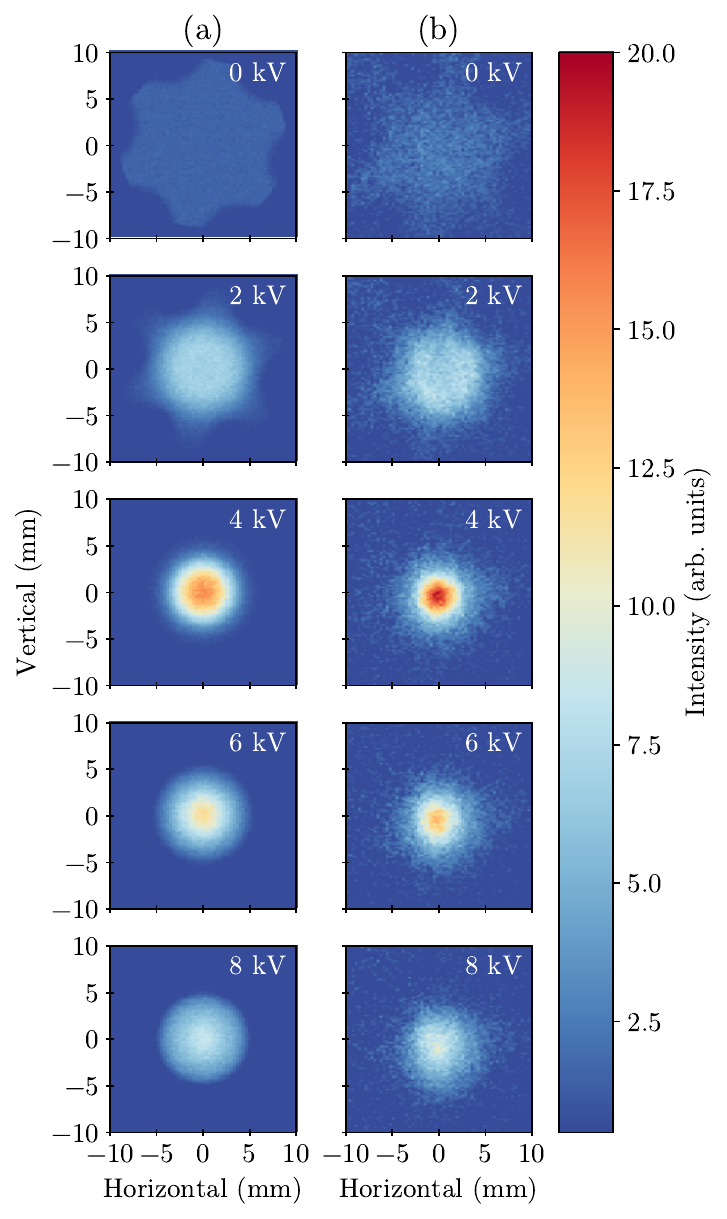}
    \caption{Transverse spread of molecular beam in the $N = 1, J = 3/2$ at the detection zone for different voltages applied to the hexapole electrodes from simulations (a) and measurements (b). All images are converted to the same color scale.}
    \label{fig:Simulation measurement comparison}
\end{figure}

\begin{figure*}[t]
    \centering
    \includegraphics[width=.9\textwidth]{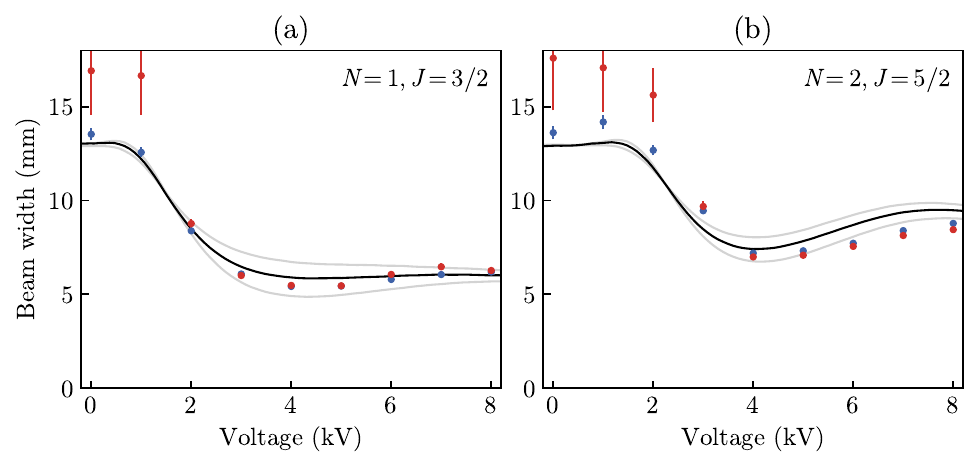}
\caption{Gaussian \textsc{fwhm} of the molecular beam at the detection zone for molecules in the $N=1, J=3/2$ (a) and $N=2, J=5/2$ (b) states as function of the voltage magnitude applied to the hexapole. The beam projection along the horizontal (blue) and vertical axis (red) are presented separately. The solid lines show the prediction from trajectory calculations, for an initial \textsc{fwhm} beam width of 6\,mm (black), 4.5 and 7.5\,mm (grey).}
    \label{fig:Beam widths}
\end{figure*}

\section{Molecular beam widths and intensities}
To examine the performance of the hexapole lens the beam characteristics have been measured at 712\,mm from the cell exit, at the position where the molecular beam will enter the traveling-wave Stark decelerator\,\cite{Aggarwal2018}. 
For these measurements, the laser frequency was kept fixed at transition 2 or 8, while the power of the laser sheet was increased to 700\,$\mu$W. At this intensity, the lines are broadened to about 25 MHz, which is sufficiently large to detect molecules independent of their transverse velocity. In this case several transitions are addressed simultaneously. For $N = 1$, transition 1-4 contribute, which implies that all 8 components of the $N = 1, J = 3/2$ are detected, of which 4 are low-field seeking ($|M_N| = 0$) and 4 are high-field seeking ($|M_N| = 1$). For $N = 2$, transition 8-12 contribute, which implies that all 12 components of the $N = 2, J = 5/2$ are detected, of which 4 are low-field seeking ($|M_N| = 0$), 4 are high-field seeking ($|M_N| = 2$) and 4 are in between ($|M_N| = 1$). \\
\indent \autoref{fig:Simulation measurement comparison} shows the effect of variation of the hexapole voltage on the molecular beam both from simulations (a) and imaging on the \textsc{emccd} camera (b). It shows an increasing focusing effect up to 4\,kV, after which the intensity starts decreasing and the beam starts to expand. As discussed in \autoref{sec:Force and Stark}, this corresponds to under- and overfocusing with a soft-focus behind or in front of the detection point, respectively. When a voltage of 2\,kV is applied, the molecular beam is collimated. At this voltage, a vague donut shape can be observed in the recorded image caused by spherical aberration of the lens. The donut is washed out by the longitudinal velocity spread in the molecular beam.\\
\indent \autoref{fig:Beam widths} shows the Gaussian \textsc{fwhm} beam width of the molecular beam projections as well as the prediction from the simulation for both the (a) $N = 1, J = 3/2$ and (b) $N = 2, J = 5/2$ states. The beam projections are found by integrating the images over the horizontal or vertical axis, corresponding to the axis along the laser sheet and orthogonal to it respectively. Examples of these projections and the corresponding Gaussian fits are shown in \autoref{fig:Axis projection}. The beam projection along the horizontal (blue data points) and vertical axis (red data points) are presented separately. The error bars represent the uncertainty in the least squares fit. As the widths along both directions are the same within the error, the method used for compensating the camera angle is validated. Furthermore, this confirms that the intensity of the laser sheet is sufficiently homogeneous and that all Doppler components of the molecular beam are detected equally. \\
\indent The solid lines also shown in \autoref{fig:Beam widths} result from trajectory calculations, using as input the longitudinal velocity distribution from the measured time-of-flight profiles shown in \autoref{fig:Time of flight} and the transverse velocity spread taken from the absorption spectrum measured directly behind the source. The initial transverse spread, corresponding to the size of the molecular beam at the so-called freezing plane, from where collisions in the molecular beam are negligible\,\cite{Pauly2000}, is a priori not known and is a free parameter. The black line is simulated using a Gaussian \textsc{fwhm} position spread of 6\,mm (black) while the grey curves are calculated using a spread of 4.5 and 7.5\,mm. As our cryogenic buffer gas source is operated in the regime between supersonic and effusive expansion, an effective beam diameter slightly larger than the 4.5\,mm cell aperture seems reasonable\,\cite{Hutzler2012}. 

\begin{figure*}[ht]
    \centering
    \includegraphics[width=.9\linewidth]{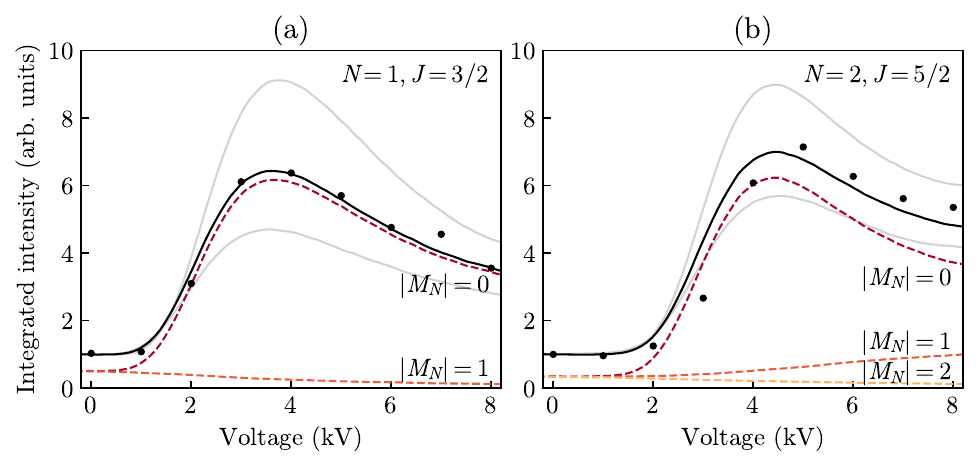}
    \caption{Intensity of a beam of molecules in (a) the $N = 1, J = 3/2$ and (b) the $N = 2, J = 5/2$ state integrated over a 4\,mm diameter circular area as function of the voltage applied to the hexapole lens. The solid lines result from trajectory calculations for an initial beam with a \textsc{fwhm} transverse position spread of 6\,mm (black), 4.5 and 7.5\,mm (grey).
    The dashed lines correspond to the contributions of the different $M_N$ components to the simulation shown in black.}
    \label{fig:Beam intensity}
\end{figure*}

\indent \autoref{fig:Beam intensity} shows the beam intensities integrated over a 4\,mm diameter circular area around the center of the molecular beam for different voltage magnitudes applied to the hexapole lens, for molecules in (a) the $N = 1, J = 3/2$ and (b) the $N = 2, J = 5/2$ states. Again the result for trajectory simulations is shown as a solid line, with the same initial beam widths at the freezing plane as in \autoref{fig:Beam widths}. The contributions of the $|M_N|$ components are simulated separately and shown for the 6\,mm initial beam width as colored dashed lines in \autoref{fig:Beam intensity}. The detected transitions probe ground states with equal contributions of each of the $|M_N|$ components, corresponding to equal contributions to the integrated intensity when the hexapole lens is off. The maximum intensity gain for molecules in the low-field seeking $N = 1, |M_N| = 0$ component is 12, reached at an applied voltage magnitude of 3.7\,kV, close to where the electric field strength at the electrodes is equal to the turning point field $E_{tp}$. \\
\indent For molecules in the low-field seeking $N = 2, |M_N| = 0$ component, a similar maximum gain factor is found, despite the $\sim$2 times larger capture velocity which should result in a $\sim$4 times larger gain factor. At the voltage at which the maximum gain is obtained, the electric field at the electrodes is well below the turning point field $E_{tp}$, for which a voltage of 11.4\,kV has to be applied. However, in this case, the focal point for this hexapole lens lies before the detection zone. A hexapole lens aimed for molecules in the $N = 2$ should have a shorter length than the current one, designed for molecules in the $N = 1$. Alternatively, the effective hexapole length could be adjusted with pulsed voltages as for example is done in\,\cite{Cheng2016}.

\begin{figure*}[ht]
    \centering
    \includegraphics[width=.9\linewidth]{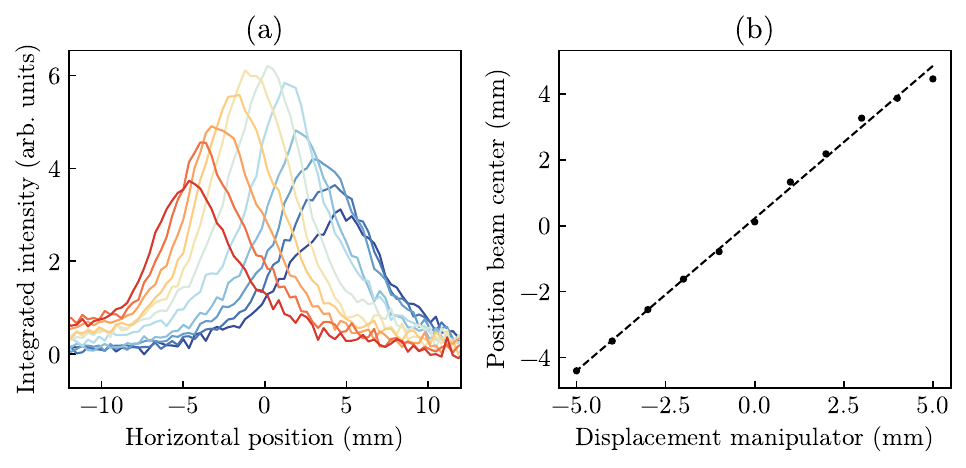}
    \caption{Horizontal translation of the molecular beam when the hexapole lens is moved using a manipulator. (a) The projection of the full beam onto the horizontal axis with the hexapole lens being displaced in steps of 1\,mm. (b) The resulting center position of the beam found from a Gaussian fit to the projections (data points). The linear fit (dashed line) shows a translation factor of 0.93(2) over a range of 10\,mm. }
    \label{fig:Beam alignment}
\end{figure*}
\section{Beam deflection and alignment}
\indent The tube that holds the hexapole lens can be moved relative to the source chamber by a XY-manipulator (Hositrad HMC1000). In this way the hexapole lens can be moved in both transverse directions separately, orthogonal to the molecular beam axis. \autoref{fig:Beam alignment}(a) shows the horizontal distribution of the molecular beam at the detection zone after traversal of the hexapole lens for different displacements of the manipulator along the horizontal axis. In these measurements a voltage of 4\,kV was applied to the hexapole. The displacement of the center of the beam is found from a Gaussian fit to these projections and is plotted in \autoref{fig:Beam alignment}(b).
As observed, the beam displacement is linear with the displacement of the manipulator, with a translation factor of 0.93(2). The beam can be displaced by several millimeter without severe intensity loss. Similar results have been obtained for the vertical displacement. The ability to move the beam can be used to fine-tune the alignment between the source and the downstream setup, such as a traveling-wave Stark decelerator.
\section{Conclusions and outlook}
We demonstrate that beams of barium fluoride molecules can be focused using a hexapole lens. State selective imaging based on laser-induced fluorescence captured on an \textsc{emccd} camera is used to record the spatial distribution of the molecules. It is shown that the number of molecules in the $N = 1, |M_N| = 0$ state reaching a 4\,mm diameter target area at a distance of 712\,mm from the source is increased by a factor of 12, limited by spherical aberrations caused by the non-linearity of the Stark curves and the low capture velocity. In addition, we demonstrate that the molecular beam can be displaced by up to $\pm$5 mm by moving the hexapole lens.\\
\indent The main motivation for this work is to increase the number of molecules that enters our traveling-wave Stark decelerator. In previous experiments\,\cite{Aggarwal2021}, there was a 370\,mm gap between the cryogenic source and the decelerator to allow for separation of the source and decelerator chambers by a vacuum valve. By free-flight propagation of the initial beam distribution, it can be calculated that, for BaF in the $N=1, |M_N| = 0$ state, this gap results in a factor of 2.6 loss in number of molecules in the transverse phase-space acceptance of the decelerator. For molecules with a larger Stark shift and lower mass, the decelerator has a larger acceptance resulting in a greater loss at the same distance. By implementing a hexapole the gap was increased to 712\,mm, resulting in a loss factor of 7.7 when no voltages are applied. From our trajectory calculations we find that, at the optimal voltage, the number of molecules that fall on the transverse phase-space of the acceptance is increased by a factor of 6.5. Consequently, with a hexapole, the loss factor is only 1.2. Or equivalently, 84$\%$ of the maximum molecular beam intensity, that would occur for zero separation between the source and decelerator, is regained by the hexapole lens. Compared to the minimal distance of 370\,mm given by practical limitations a factor of 2.5 is gained.\\
Increasing the distance between the beam source and decelerator from 370\,mm to 720\,mm results in an increase of the longitudinal phase-space matching by 5$\%$ in experiments where molecules are decelerated from 200\,m/s to 30\,m/s in 4.5\,m. For a more significant increase, the distance would need to be increased to several meters\,\cite{Fabrikant2014}. \\
\indent The level structure of BaF allows for laser-cooling\,\cite{Hao2019, Zhang2022, Fitch2021}. An exciting prospect is to combine electrostatic lenses, that provide a force that depends only on the transverse position of the molecules, with laser-cooling techniques, that provide a force that depends only on the transverse velocity of the molecules. In this way, beams are both cooled and focused. The resulting dense and cold beam of molecules can be used for precision tests directly, or as input for a Stark decelerator.\\  

\section*{Acknowledgment}
 The NL-\textit{e}EDM consortium receives program funding (\textit{e}EDM-166 and XL21.074) from the Netherlands Research Council (NWO). We thank prof. S Stolte for kindly providing us with the hexapole lens. We acknowledge the technical support from R Kortekaas, L Huisman and O B\"oll.

\bibliography{ps_mendeley.bib} 
\end{document}